\documentclass{iaus}
\usepackage{graphicx}

\title[Magnetic fields in in $\beta$\,Cephei and SPB stars] 
{New magnetic field measurements 
of $\beta$\,Cephei stars and Slowly Pulsating B stars}

\author[S. Hubrig et al.]   
{S.~Hubrig$^1$,
M. Briquet$^2$, P. De Cat$^3$, M.~Sch\"oller$^4$, T. Morel$^5$ \and I.~Ilyin$^6$}

\affiliation{$^1$ESO, Santiago, Chile;
email: shubrig@eso.org;
$^2$Katholieke Universiteit Leuven, Belgium;
$^3$Koninklijke Sterrenwacht van Belgi\"e, Brussel, Belgium;
$^4$ESO, Garching, Germany;
$^5$Universit\'e de Li\`ege, Belgium;
$^6$AIP Potsdam, Germany}

\pubyear{2008}
\volume{259}  
\pagerange{1--2}
\date{?? and in revised form ??}
 \jname{Cosmic Magnetic Fields: from Planets,
to Stars and Galaxies } \editors{K.G. Strassmeier, A.G. Kosovichev
\& J. Beckmann, eds.}
\begin{document}

\maketitle

\begin{abstract}
We present the results of the continuation of our magnetic survey with FORS\,1 at the VLT of a sample of 
B-type stars consisting of confirmed or candidate $\beta$\,Cephei stars and Slowly Pulsating B  
stars.
Roughly one third of the studied $\beta$\,Cephei stars have detected magnetic fields. The fraction of magnetic 
Slowly Pulsating B and candidate Slowly Pulsating B stars is found to be higher, up to 50\%. We find that 
the domains of
magnetic and non-magnetic pulsating stars in the H-R diagram largely overlap, and 
no clear picture emerges as to the possible evolution of the magnetic field across the main sequence.

\keywords{stars: variables: $\beta$\,Cephei, stars: variables: SPBs, stars: oscillations, stars: Hertzsprung-Russell diagram, 
stars: magnetic fields, techniques: polarimetric, stars: individual ($\xi^1$~CMa)}
\end{abstract}

\firstsection 
\section{Introduction}
In our first publication on the magnetic survey of pulsating B-type stars
(\cite[Hubrig \etal{} 2006]{Hubrig2006}), 
we announced detections of weak mean longitudinal 
magnetic fields of the order of a few hundred~Gauss in 13~Slowly Pulsating B (hereafter SPB) stars and in 
the $\beta$~Cephei star $\xi^1$~CMa.
$\xi^1$\,CMa showed a mean longitudinal magnetic field of the order of 300\,G.
Since the role of magnetic fields in modeling oscillations of B-type stars remained to be studied,
we collected additional 98 magnetic field measurements in a sample of 60 pulsating and candidate pulsating stars.

\section{Observations and analysis}\label{sec:observ}
The spectropolarimetric observations have been carried out in the 
years 2006 to 2008 at the
European Southern Observatory with FORS\,1 mounted on the 8-m Kueyen telescope of the VLT.
For the major part of the observations we used the GRISM\,600B in the wavelength range 3480--5890\,\AA{}
to cover all hydrogen Balmer lines from H$\beta$ to the Balmer jump. One additional 
high resolution observation of $\xi^1$\,CMa has been obtained with the SOFIN echelle spectrograph 
installed at the 2.56\,m Nordic Optical Telescope. The Stokes~$I$ and $V$ spectra are 
presented in the left panel of Fig.~1. In the right panel of 
Fig.~1 we present magnetic field measurements of the same star using FORS\,1 over the last 4.4\,years.
Among the SPBs and $\beta$\,Cephei stars, the detected magnetic fields are mainly of the order of 100--200\,G. 

\section{Discussion}
Roughly one third of the $\beta$\,Cephei stars have detected magnetic fields: Out of 13 $\beta$\,Cephei stars 
studied to date with FORS\,1, four stars possess weak magnetic fields, 
and out of the sample of six suspected $\beta$\,Cephei stars two show a weak magnetic field. 
The fraction of magnetic SPBs and candidate SPBs is found to be higher: roughly half of the 34 SPB 
stars have been found to be magnetic and among the 16 candidate SPBs, eight stars possess magnetic fields.

About a dozen pulsating stars discussed in our previous study \cite{Hubrig2006} have been subsequently
observed by Silvester et al.\ (this proceedings) with ESPaDOnS at CFHT and NARVAL at TBL.
The interesting result is that they could confirm our previous non-detections in ten  $\beta$\,Cephei and SPB 
stars but failed to detect weak magnetic fields in the other four magnetic SPB stars selected from our study.
On the other hand, as has been shown at the present meeting by Henrichs et al.\ 
(this proceedings) who confirmed our detection of a magnetic field in the B3V star 16\,Peg, 
the integrated Zeeman features obtained from the application of a cross-correlation method,
the Least-Squares Deconvolution,  are extremely weak for 
$\sim$100\,G fields. This means, that even using high-resolution spectrographs like ESPaDOnS or Narval,
very high signal-to-noise observations are mandatory to be able to detect magnetic fields in hot stars where 
only a limited number of metal lines are suitable for the measurements. 

In an attempt to understand why only a fraction of the pulsating stars exhibit magnetic fields, we
studied the position of magnetic and non-magnetic pulsating stars in the H-R diagram. 
We find that their domains in the H-R diagram largely overlap, and 
no clear picture emerges as to the possible evolution of the magnetic field across the main sequence.
It is possible that stronger fields tend to be found in stars with lower pulsating frequencies and 
smaller pulsating amplitudes. A somewhat similar trend is found if we consider a correlation
between the field strength and the $v$\,sin\,$i$-values, i.e.\
stronger magnetic fields tend to be found in more slowly rotating stars.

\begin{figure}
\centering
\includegraphics[height=1.8in,angle=0]{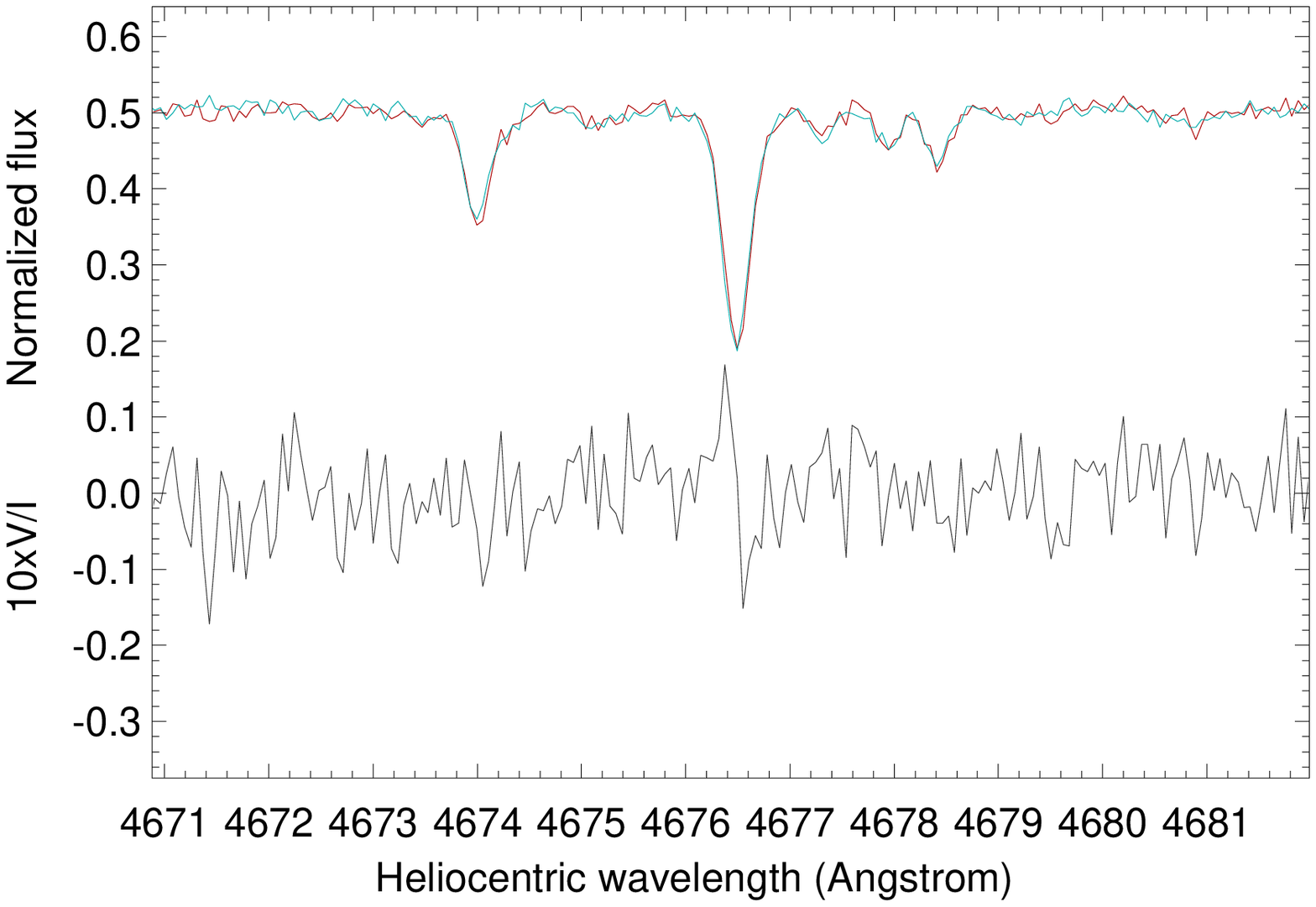}
\includegraphics[height=1.8in,angle=0]{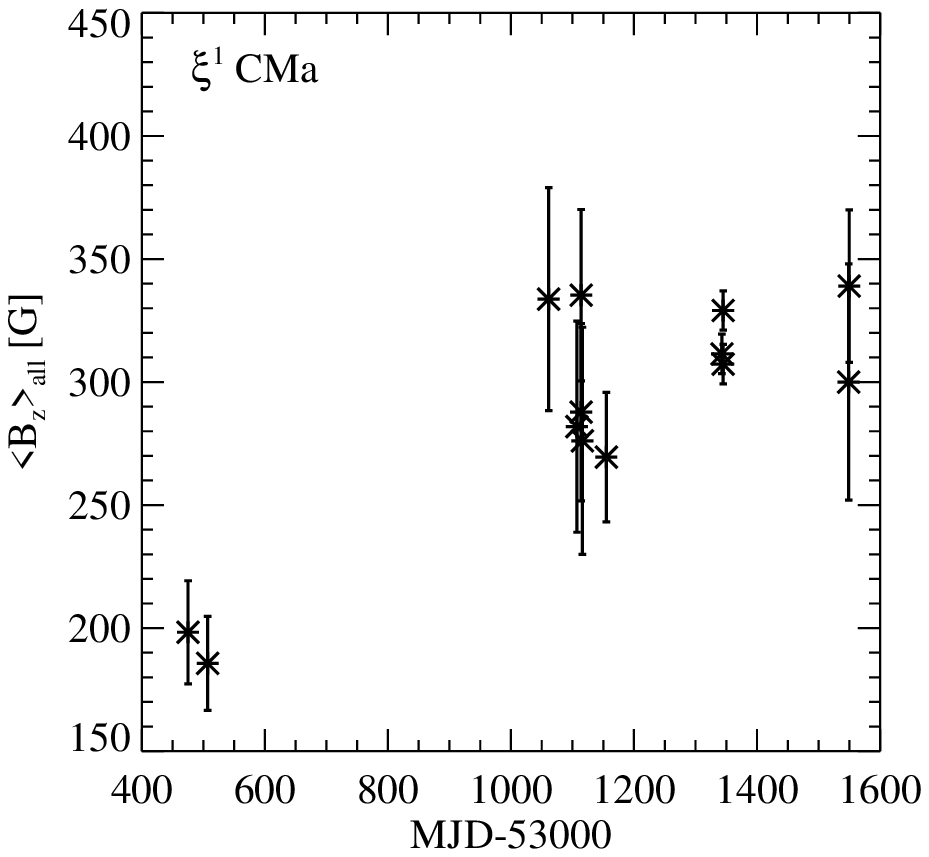} 
\caption{Left panel: High-resolution (R$\approx$30\,000) SOFIN polarimetric spectrum of $\xi^1$\,CMa.
A clear Zeeman feature is detected at the position of the unblended line O~II $\lambda$4676.2.
Right panel: Magnetic field measurements with FORS\,1 of $\xi^1$\,CMa over the last 4.4~years
 } \label{fig:Ca}
\end{figure}


%


\end{document}